\documentclass[twocolumn,fleqn,10pt]{wlscirep}
\usepackage[
    backend=biber,
    natbib=true,
    style=phys,
    maxbibnames=8,
    date=year,
    sorting=none
]{biblatex}
\usepackage[utf8]{inputenc}
\usepackage[T1]{fontenc}
\usepackage[normalem]{ulem}
\usepackage{gensymb}

\usepackage{todonotes} 
\usepackage{lineno}
\usepackage{float}
\usepackage{dblfloatfix}

\renewcommand{\thesection}{\Roman{section}}
\renewcommand{\thesubsection}{\Alph{subsection}}
\renewcommand{\thesubsubsection}{\arabic{subsubsection}}

\title{Nanoscale Three-Dimensional Imaging of Integrated Circuits using a Scanning Electron Microscope and Transition-Edge Sensor Spectrometer}

\author[1,2a)]{Nathan Nakamura}
\author[1,2]{Paul Szypryt}
\author[3]{Amber L. Dagel}
\author[1]{Bradley K. Alpert}
\author[1]{Douglas A. Bennett}
\author[1]{W. Bertrand Doriese}
\author[1,2]{Malcolm Durkin}
\author[1,2]{Joseph W. Fowler}
\author[3]{Dylan T. Fox}
\author[1,2]{Johnathon D. Gard}
\author[3]{Ryan N. Goodner}
\author[3]{J. Zachariah Harris}
\author[1]{Gene C. Hilton}
\author[3]{Edward S. Jimenez}
\author[3]{Burke L. Kernen}
\author[3]{Kurt W. Larson}
\author[4]{Zachary H. Levine}
\author[3]{Daniel McArthur}
\author[1,2]{Kelsey M. Morgan}
\author[1]{Galen C. O'Neil}
\author[1,2]{Nathan J. Ortiz}
\author[1,2]{Christine G. Pappas}
\author[1]{Carl D. Reintsema}
\author[1]{Daniel R. Schmidt}
\author[3]{Peter A. Schultz}
\author[3]{Kyle R. Thompson}
\author[1,2]{Joel N. Ullom}
\author[1]{Leila Vale}
\author[3]{Courtenay T. Vaughan}
\author[3]{Christopher Walker}
\author[1,2]{Joel C. Weber}
\author[3]{Jason W. Wheeler}
\author[1]{Daniel S. Swetz}

\affil[1]{National Institute of Standards and Technology, Boulder, Colorado 80305, USA}
\affil[2]{Department of Physics, University of Colorado, Boulder, Colorado 80309, USA}
\affil[3]{Sandia National Laboratories, Albuquerque, New Mexico 87123, USA}
\affil[4]{National Institute of Standards and Technology, Gaithersburg, Maryland 20899, USA}
\affil[a)]{nathan.nakamura@nist.gov}

\begin{abstract}
X-ray nanotomography is a powerful tool for the characterization of nanoscale materials and structures, but is difficult to implement due to competing requirements on X-ray flux and spot size. Due to this constraint, state-of-the-art nanotomography is predominantly performed at large synchrotron facilities. We present a laboratory-scale nanotomography instrument that achieves nanoscale spatial resolution while changing the limitations of conventional tomography tools. The instrument combines the electron beam of a scanning electron microscope (SEM) with the precise, broadband X-ray detection of a superconducting transition-edge sensor (TES) microcalorimeter. The electron beam generates a highly focused X-ray spot in a metal target held micrometers away from the sample of interest, while the TES spectrometer isolates target photons with high signal-to-noise. This combination of a focused X-ray spot, energy-resolved X-ray detection, and unique system geometry enable nanoscale, element-specific X-ray imaging in a compact footprint. The proof-of-concept for this approach to X-ray nanotomography is demonstrated by imaging 160~nm features in three dimensions in 6 layers of a Cu-SiO$_2$ integrated circuit, and a path towards finer resolution and enhanced imaging capabilities is discussed.

\end{abstract}
\begin{document}
\maketitle
\section{Introduction}

Nanoscale three-dimensional X-ray imaging yields insight into the internal structure of materials which would otherwise be difficult to probe, with applications spanning many emerging technologies such as advanced energy storage materials\supercite{nanotomo_energy} and next-generation integrated circuits (ICs).\supercite{IC_imaging} Modern nanoelectronics in particular would benefit from advanced nanoscale imaging, with recent reports calling out nanoscale resolution and sensitivity to material composition as potentially transformative capabilities for nanoelectronics characterization.\supercite{ITRS} X-ray absorption computed tomography (CT) is a well-developed method for three-dimensional imaging, relying on differences in X-ray transmission to determine the size and shape of subsurface features.\supercite{Withers_genTomo} A CT scan consists of acquiring many two-dimensional radiographs of a sample at different projection angles and combining these projections into a three-dimensional reconstructed image. In general, the X-ray focal spot size must be comparable to the desired spatial resolution when collecting each radiograph. This requirement limits the number of available photons when pushing the spatial resolution to the order of tens to hundreds of nanometers, leading to long collection times and low signal-to-noise. For this reason, X-ray nanotomography has largely been deployed at synchrotron beamlines, where a high X-ray flux can be maintained even with nanoscale X-ray spot sizes.\supercite{ssrl_nanotomo_2020, synch_nanotomo_book, Attwood_2010}

\begin{figure*}[!t]
\centering
\includegraphics[width=\textwidth]{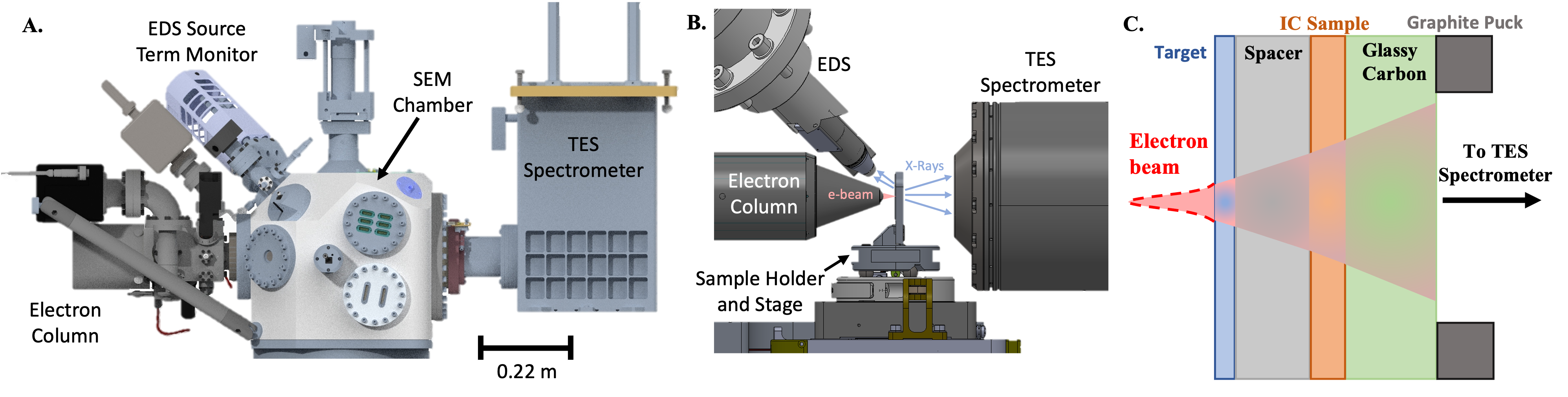}
\caption{(A)~MINT overview, consisting of an electron column, energy dispersive spectroscopy (EDS) source term monitor, SEM chamber, and TES spectrometer. (B)~View inside the SEM chamber, showing the electron beam incident on a sample in the sample holder and generated X-rays going to the TES and EDS. (C)~Schematic demonstrating the MINT sample configuration and x-ray generation in sample layers: an electron beam incident on a target layer generates X-rays in a nanoscale spot size, which are attenuated by the IC and detected by the TES. Electrons not stopped in the target layer spread into a larger spot size and generate X-rays in other layers of the sample. Sample thicknesses are not drawn to scale.
Part B is reprinted from Ref.~\cite{weichman2020fluorescent} with permission (http://creativecommons.org/licenses/by/4.0/).}
\label{fig:overview}
\end{figure*}

A need exists for both more accessible and disseminable nanotomography systems, capable of resolving buried nanoscale features in modern ICs in a compact instrument footprint. Despite this need, current tomography instruments do not provide the capabilities required for nanoscale IC characterization. Current laboratory-scale tools capable of achieving the best spatial resolution rely on X-ray optics and destructive sample preparation techniques.\supercite{Tan_convTomo_2020, muller_convTomo_2020} The highest spatial resolution is achieved at lower X-ray energies, which leads to high attenuation and low detected X-ray flux in thicker samples or high-Z components. This results in the majority of laboratory nanotomography instruments destroying the sample outside of a pre-selected micrometer-scale region, severely limiting the available scan area. This is a key limiting factor in nanoelectronics inspection, as the location of compromised features is often unknown prior to imaging and thus the region-of-interest cannot be reliably chosen prior to performing tomography. Another approach to achieving high-resolution X-ray CT in a compact footprint is to utilize the electron beam from a scanning electron microscope (SEM) and a metal target as the X-ray source. A small number of SEM-based systems have been developed,\supercite{Mayo_SEMtomo_2002, Perini_SEMtomo_2017, Pauwels_SEMtomo_2010, muller2021novel}, but none reach the spatial resolution requirements to characterize features smaller than 200~nm. Additionally, current tools characterize only the size and shape of sample features, and provide limited information regarding the elemental composition. Alternative approaches utilizing novel X-ray target design, X-ray detection, and system geometry are still required to improve the achievable spatial resolution and imaging capabilities of laboratory CT systems.

Here, we demonstrate MINT (Microscope for Integrated circuit NanoTomography), a compact X-ray CT system capable of achieving nanoscale spatial resolution and elemental sensitivity in 3-dimensions. MINT is governed by a different set of limitations than conventional approaches to X-ray nanotomography, allowing it to image nanoscale features without the use of X-ray focusing optics, while preserving large area planar samples, and with information regarding the elemental composition of samples. MINT combines the electron beam of a scanning electron microscope (SEM) with a transition-edge sensor (TES) X-ray spectrometer (Fig. \ref{fig:overview}A-B). A highly focused electron beam incident on a thin-film metal target generates X-rays, which are attenuated by the sample and detected by the TES spectrometer (Fig. \ref{fig:overview}C). This results in an instrument with a nanoscale X-ray spot size, efficient and energy-resolved X-ray detection, and system geometry enabling high magnification. Electron beam spots can routinely be focused to the nanometer scale, enabling a nanoscale X-ray generation volume within the thin-film target. The X-rays generated in this nanoscale spot consist of both characteristic fluorescence lines at a specific energy and a broadband bremsstrahlung background. Additional bremsstrahlung will be generated from other layers in the sample stack with a more diffuse spot size, due to electrons which are not stopped in the target and produce X-rays in subsequent layers. To maintain nanoscale spatial resolution, only photons generated in the nanoscale spot size in the metal target layer should be used for tomographic reconstruction. Therefore, an energy-resolving X-ray detector is required to distinguish characteristic target X-rays from background photons generated elsewhere in the sample. MINT utilizes a TES spectrometer for X-ray detection, which consists of hundreds of superconducting microcalorimeter TES pixels.\supercite{Doriese_TES_beamline_2017} A TES microcalorimeter pixel is an efficient single-photon counting detector operated at cryogenic temperatures, capable of extremely high energy resolving power and detection across a broadband energy range with high precision.\supercite{ullom_review_2015} Each TES pixel can thus isolate characteristic X-rays generated in the target layer with high signal-to-noise, ensuring that only X-rays generated in the nanoscale target generation volume are used for tomographic reconstruction. Energy resolved detection also enables energy binning in reconstructions, which can enable elemental specificity and can improve the final image contrast.\supercite{McCollough_dualCT_2015, Shikhaliev_dualCT_2008} The use of a thin-film metal target also allows the target layer to be deposited directly onto the sample, which generates the X-ray source within micrometers of the sample. This allows for high geometric magnification while keeping the detector relatively close to the source, improving the solid angle coverage and X-ray flux at the detector plane. This combination of a highly focused electron beam, thin-film metal target, efficient, energy-resolved X-ray detection, and high geometric magnification enables nanoscale three-dimensional imaging in a compact instrument. The total footprint of MINT is 1.65~m $\times$ 1.30~m and it is 2.00~m tall at its highest point, compact enough to operate in most standard laboratory spaces. This approach to laboratory X-ray tomography provides an alternative method for nanoscale imaging with a promising path forward.

MINT was developed with nanoelectronics as the application space, and so an IC consisting of Cu wiring in an SiO$_2$ dielectric fabricated at the 130~nm technology node is used to demonstrate imaging capability. ICs are a critical use-case for nanotomography, as modern ICs are comprised of nanoscale features and many fabrication layers. This structural complexity makes precise imaging and characterization of subsurface layers difficult, limiting critical IC diagnostics such as defect detection and failure analysis.\supercite{IC_imaging, Mahmood_IC_2015} ICs pose an additional challenge for traditional sample preparation, as their planar geometry either inhibits rotation or requires that all but a small subsection of the IC be destroyed. This eliminates the ability to scan multiple regions of a sample and limits the ability use the same sample for complementary diagnostic measurements. The source, sample, and system geometry, along with a custom tomographic reconstruction algorithm, allows MINT to overcome this challenge and keep planar samples intact during imaging.

MINT is the first implementation of a TES X-ray spectrometer for three-dimensional CT imaging and represents a proof-of-concept for the combination of an SEM, TES, and the novel sample and system geometry presented here. The current instrument is photon-limited, and it is expected that higher efficiency X-ray detection or higher pixel-count TES spectrometers will be required to achieve faster imaging speeds. Our prior work has described the development of a higher pixel-count spectrometer for three-dimensional imaging applications, but the development of this spectrometer is outside the scope of the current work.\supercite{Szypryt_1k_2023, Szypryt_3k_2021} Prior work has also described the tomographic reconstruction approach required to generate three-dimensional images from MINT data, but does not provide details on the system physics or instrumental approach to enable laboratory nanoscale CT of integrated circuits.\supercite{levine2023tabletop} Here, the first description of the instrument design and underlying physics necessary to achieve nanoscale three-dimensional X-ray imaging with MINT is presented, providing a roadmap for nanoscale X-ray CT in an SEM-based tabletop instrument. In MINT, the electron beam conditions, X-ray target materials and dimensions, system magnification, and energy of photons used in the reconstruction can be tuned for optimized imaging of a wide variety of samples and material compositions. The imaging demonstration discussed here is comparable to current state-of-the-art laboratory-scale CT instruments in terms of the achievable spatial resolution on IC samples, with additional elemental detection capabilities not found elsewhere. The unique source and system setup of MINT leads to a different set of limiting factors compared to conventional approaches to laboratory CT, and a clear path forward exists to improve the imaging speed, spatial resolution, and spectral imaging capabilities.


\section{MINT System Description} \label{sec:systemDesc}

\subsection{X-ray Source}
The X-ray source has a significant influence on the final image quality. The focal spot size of the X-ray source impacts the achieved spatial resolution,\supercite{Chen_tomoRes, Hu_tomoRes} and attenuation in the sample is a function of the incident X-ray energy as well as the size, shape, and composition of materials. In MINT, the X-ray source is a careful optimization of target and electron beam parameters to maximize imaging speed for a given sample. Ideally, the source would maximize the X-ray flux at X-ray energies which provide high imaging contrast for a given sample while maintaining a spot size near the spatial resolution goal. As the IC sample used here was fabricated at the 130~nm node, we aimed to maintain a focal spot size at or below 130~nm.

\subsubsection{Target Selection} \label{sec:targetSelection}
The thin-film target thickness and material can be selected to maximize imaging speed for given sample and desired spatial resolution. The final imaging speed will be dependent on the electron-stopping power of the target material and the energy of the characteristic X-ray lines emitted. Thicker, higher $Z$ targets will stop more electrons and thus generate more X-ray flux, but will also result in larger X-ray focal spot sizes. The energy of the emitted fluorescence line will influence the expected contrast, which we define as the difference in attenuation between Cu and SiO$_2$ at a given X-ray energy ($\Delta\mu$). Thus, a target material should be chosen to maximize X-ray generation at an energy of high contrast for the sample of interest, while maintaining a suitably small X-ray spot size for nanoscale spatial resolution.

A target was selected for the sample IC by developing an estimate of the relative imaging speed based on prospective target materials and thicknesses and the available operating conditions of the SEM electron beam. Prospective target materials were chosen to cover a range of $Z$, with fluorescence lines within the dynamic range of the TES spectrometer, and based on fabrication capabilities. The imaging speed estimate is simplified by supposing that we count photons over an energy interval of width $\Delta E$, approximately equal to the TES energy resolution, from which an expected background must be subtracted. The final prediction for the relative imaging speed is given as
\begin{equation}
    \frac{1}{t} < f \left(\Delta\mu\right)^2 (1+2b\Delta E/f)^{-1}.
    \label{equation:imaging_speed}
\end{equation}
where $f$ is the fluorescence photon rate, $b$ is the bremsstrahlung rate, and $1/t$ is the imaging speed. A full derivation of Equation \ref{equation:imaging_speed} and a method to predict the absolute imaging speed can be found in the Supplemental Material (SM), Section S1. The first two factors show that the speed scales as the fluorescence photon rate $f$ and as the square of the material absorption contrast. The last factor is a slowdown due to subtracting a background from the observed counts.

\begin{figure}[!t]
\centering
\includegraphics[width=\linewidth]{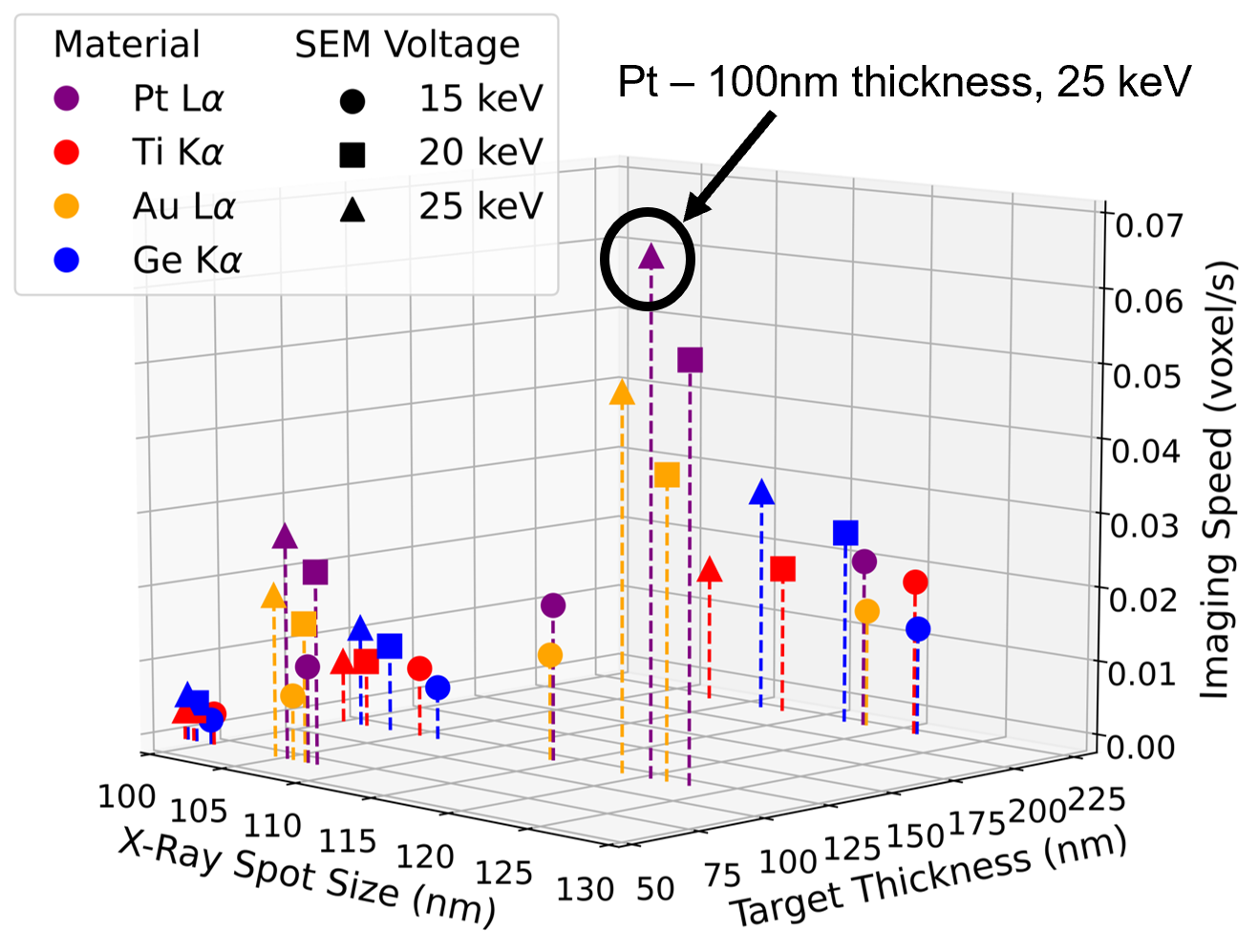}
\caption{PENELOPE simulation results from candidate target materials showing the predicted imaging speed for the coadded TES array. A 100~nm thick Pt layer at an electron accelerating voltage of 25~keV yields the best predicted imaging speed on the selected IC and thus was chosen for the MINT imaging demonstration.}
\label{fig:penelope}
\end{figure}

The photon yields \emph{f} and \emph{b} were predicted using the Monte Carlo electron-photon transport code PENELOPE (Penetration and ENErgy LOss of Positrons and Electrons), version 2018.\supercite{penelope} The library was driven via the main program PENEPMA\supercite{llovet_penepma_2017}. All PENELOPE results were scaled for the TES spectrometer collection efficiency and solid angle, and 10~nA of electron beam current. A 150~nm voxel size and 100~nm full-width half-maximum (FWHM) Gaussian electron beam spot was assumed. The true electron beam FWHM was later characterized to be slightly narrower than the beam size used in PENELOPE (See Section II.A.2). The 100~nm FWHM provides margin for error in the target thickness selection and ensures that the X-ray spot size remains well below the desired 130~nm FWHM threshold. Target materials were chosen based on which target material and thickness yielded the fastest estimated imaging speed without exceeding the desired maximum X-ray spot size.

Figure \ref{fig:penelope} shows the predicted imaging speed versus X-ray spot FWHM and target thickness for four candidate target materials at three candidate accelerating voltages. Target thicknesses or materials which exceeded the set 130~nm spot size threshold were not included. Candidate materials were selected based on fabrication capabilities and to cover a suitably wide range of X-ray fluorescence energies as to make an informed target material decision. A 100~nm thick Pt target at a 25~keV accelerating voltage yields the fastest predicted imaging speed and was chosen as the target material for this demonstration. In general, higher electron accelerating voltages produce a smaller volume of X-ray generation, allowing for thicker targets while maintaining a smaller spot size. At these higher accelerating voltages, higher $Z$ materials generate X-rays more efficiently. Additionally, the Pt L$_\alpha$ line is just above the Cu K-edge, leading to a high Cu-SiO$_2$ attenuation difference. The Pt L$_\alpha$ line is also at an energy (9.4~keV) of high TES detection efficiency, improving the number of detected photons. Lastly, higher energy fluorescence lines transmit through the silicon and glassy carbon layers with less attenuation, leading to more detected photons at the TES spectrometer. The combination of these factors leads to the 100~nm Pt target being a well-suited target for the current imaging demonstration. This target selection process can be repeated for any sample composition or imaging spatial resolution goal, allowing for flexible target design to image samples with varying composition or feature sizes.

\subsubsection{Electron Aperture Selection} \label{sec:spotsize}
\begin{figure}[!t]
\centering
\includegraphics[scale=0.75]{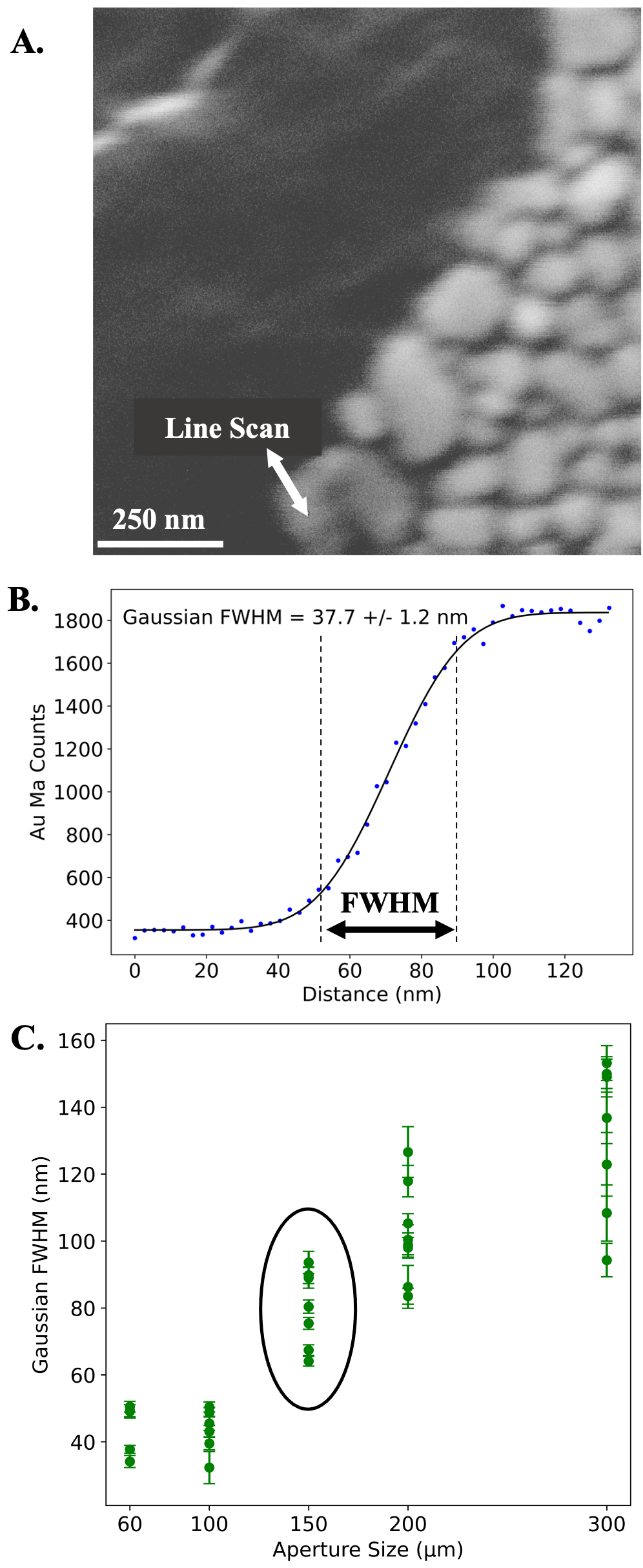}
\caption{(A)~The Au on C sample. Line scans over the edge of Au grains were collected and the Au M$_\alpha$ counts detected by the EDS were extracted. (B)~Au M$_\alpha$ counts versus distance along a line scan (blue points) using the 60~$\mu$m aperture, fit to the integral of a Gaussian (black line) to estimate the electron beam full-width half-maximum (FWHM). (C)~Estimated Gaussian FWHM versus the SEM aperture size. An aperture size of 150~$\mu$m was chosen for tomography in this measurement. At smaller aperture sizes (60~$\mu$m and below), the measured spot size becomes limited by the sharpness of the Au edge, rather than the size of the electron beam.}
\label{fig:beam}
\end{figure}

MINT utilizes a commercial horizontal electron column (Orsay Physics) with a ZrO$_2$-coated W emitter. The electron column is critical to achieving high spatial resolution, as it contributes to the source size and stability. The electron beam spot size sets the X-ray focal spot size, with a smaller electron beam on the metal target producing a smaller X-ray generation volume. The electron beam current sets the X-ray flux and thus the imaging speed, with higher beam currents producing more source X-rays.

The electron beam spot size and current are impacted by a variety of factors, including the accelerating voltage, voltages applied to the condenser and objective lenses, and the choice of aperture size. The accelerating voltage was chosen to be 25~keV, to maximize the imaging speed when using the Pt target. The condenser lens voltage was pre-selected based on electron column conditions and the objective lens voltage was adjusted at each aperture size to focus the electron beam onto the target surface. This leaves the aperture size as the last remaining variable to tune the electron beam spot size and beam current. A tradeoff exists when selecting the aperture size, as smaller aperture sizes correspond to smaller electron beam sizes but also limit the electron beam current. The largest possible aperture without exceeding the desired spot size should be chosen to maximize the electron beam current, and thus the imaging speed, at a desired spatial resolution.

To characterize the relationship between aperture size and electron beam spot size, an Au on C calibration sample was used (Ted Pella, Inc.). This sample consists of a number of Au particles distributed on a C background. The electron beam was scanned normal to Au edges (Fig.~\ref{fig:beam}A), starting on the C background and moving onto the Au particle. Multiple line scans were taken across different Au edges to average out the effect of imperfect edge sharpness on the Au particles. An energy dispersive spectroscopy (EDS) detector located on the electron beam-side of the sample and associated software (Oxford Instruments) were used to select the Au M$_\alpha$ counts. This data was then fit to the integral of a Gaussian to extract the estimated electron beam FWHM (Fig.~\ref{fig:beam}B). This process of scanning over a sharp edge has been previously demonstrated as a viable approach to estimate electron beam point spread functions.\supercite{Rishton_eSpot_1984, Gavrilenko_eSpot_2008, Goldenshtein_eSpot_1998} The measurement was repeated across five aperture sizes ranging from 60~$\mu$m to 200~$\mu$m (Fig.~\ref{fig:beam}C). The 150~$\mu$m aperture size was chosen for the current demonstration, as it yields approximately 9~nA of beam current at a Gaussian FWHM of 80~nm. Based on PENELOPE simulations, the electron beam is expected to spread by 23.6\% throughout the 100~nm thick Pt target.  This results in a predicted X-ray spot size of approximately 100~nm, safely below the desired 130~nm limit.
 
\subsection{System Geometry and Magnification}
\begin{figure}[!t]
\centering
\includegraphics[width=\linewidth]{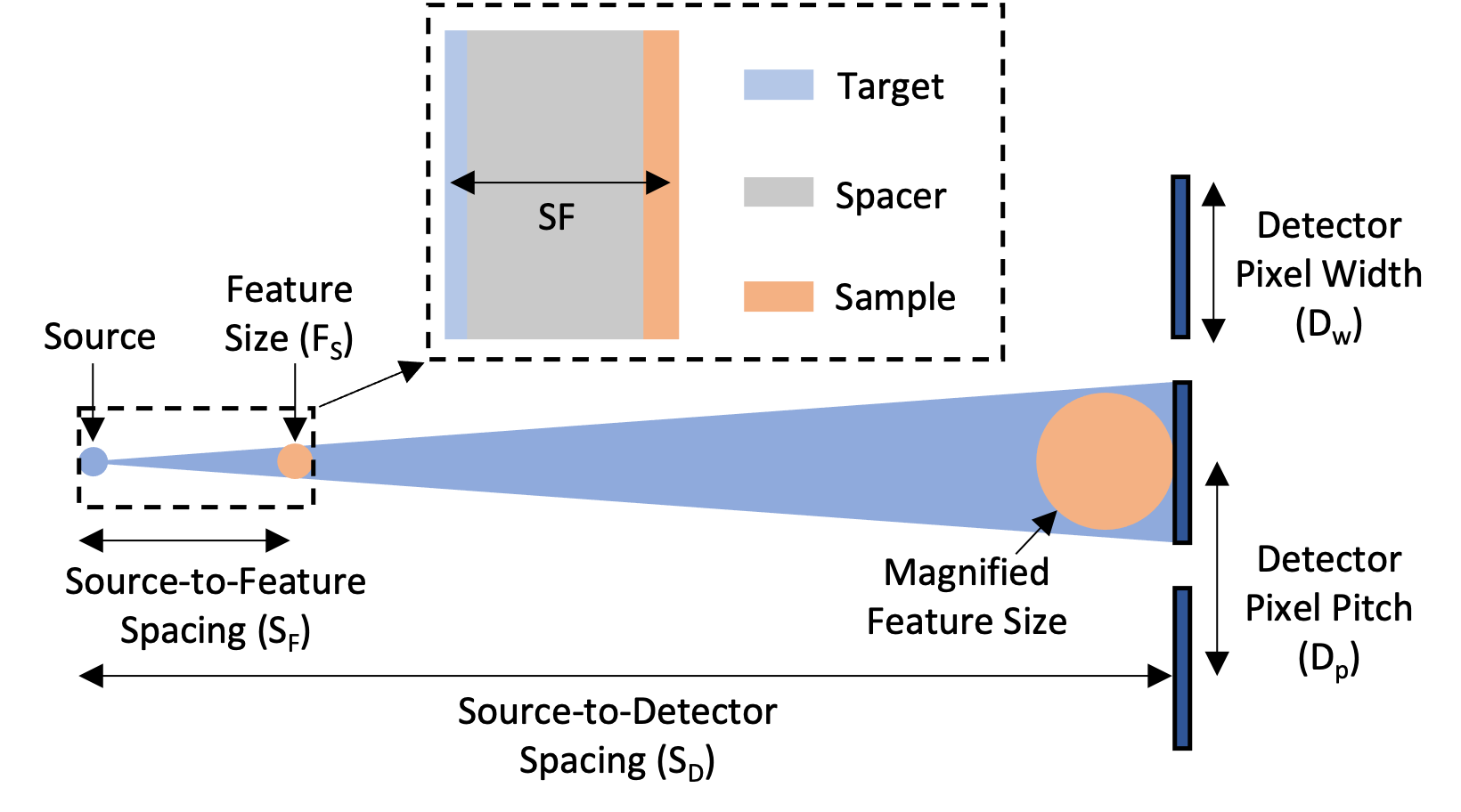}
\caption{Geometric magnification in MINT.  Nanometer-scale features in the IC are magnified onto TES pixels. The design magnification is proportional to the TES pixel pitch (D$_p$) and the desired resolvable feature size ($F_S$), while the system magnification is proportional to the ratio of the source-to-detector ($S_D$) and source-to-feature ($S_F$) spacing. The system magnification should be higher than the design magnification to resolve the desired feature size.
This figure is reprinted from Ref.~\cite{levine2023tabletop}.}
\label{fig:magnification}
\end{figure}

The geometric magnification of an X-ray CT instrument is one factor determining the achievable spatial resolution (Fig. \ref{fig:magnification}). There are two magnification factors to consider: the projection magnification \emph{M}$_{projection}$ and the system magnification \emph{M}$_{system}$. \emph{M}$_{projection}$ is the magnification of a certain feature size onto the X-ray detector, and is given by
\begin{equation}
M_{\rm projection} = 2 \frac{D_p}{F_S}
\end{equation}
where \emph{D$_p$} is the TES pixel pitch and \emph{F$_S$} is the minimum desired feature size. In the current TES spectrometer, each TES pixel has an aperture 320~$\mu$m wide, setting the pixel width. Since pixels in an array do not touch and the pixel pitch (550~$\mu$m) is larger than the pixel width, \emph{D$_p$} must be used to determine the projection magnification. A factor of 2 was included and used to design the imaging geometry to provide margin for error, as explained further below. The minimum expected feature size in the demonstration IC sample was expected to be 160~nm, leading to a projection magnification of 6875.

To resolve a given feature size the system magnification should equal or exceed the projection magnification. \emph{M}$_{system}$ is defined as the ratio of the source-to-detector distance (\emph{S$_D$}) and the source-to-feature distance (\emph{S$_F$})
\begin{equation}
M_{\rm system} = \frac{S_D}{S_F}
\end{equation}
High system magnification can be achieved either by moving the detector far from the source or by moving the source close to the sample. In practice, moving the X-ray detector far from the source reduces the solid angle of collection and X-ray flux, leading to increased imaging times and reduced signal-to-noise. In MINT, the detector is moved as close to the source as possible without collisions in the chamber, to a distance of 75~mm. High \emph{M}$_{system}$ is still achieved by depositing the X-ray target directly onto the IC wafer, resulting in a small \emph{S$_F$} (SM, Section S2). \emph{S$_F$} is set by the target thickness, the sample thickness, and the thickness of a silicon layer left between the sample and target, referred to as the spacer layer. The thickness of the spacer layer can be tuned to achieve a specific \emph{M}$_{system}$ based on the desired resolvable feature size and resulting \emph{M}$_{projection}$. In this demonstration, the spacer was thinned to 8.5~$\mu$m, which with a target thickness of 100~nm and an approximate IC sample thickness of 3.5~$\mu$m results in a system magnification of ~7300 at normal incidence. This is larger than the projection magnification, indicating that the system geometry of MINT was designed to image features down to at least 160~nm.

\emph{M}$_{system}$ will vary with the angle of incidence and depth in the sample, and so is used as an estimator to ensure that the system geometry is compatible with the desired spatial resolution. As the angle of incidence $\theta$ changes, the source-to-feature distance increases due to the effective thickness of the spacer scaling as $\sec\theta$. This reduces \emph{M}$_{system}$ at larger angles. The factor of 2 was included in the design magnification to account for this effect. In general, system parameters such as the TES pixel size and pitch, the source-to-detector distance, and the spacer thickness can be varied to achieve magnification factors suitable for a desired feature size.


\subsection{TES Spectrometer}
\begin{figure}[!t]
\centering
\includegraphics[width=\linewidth]{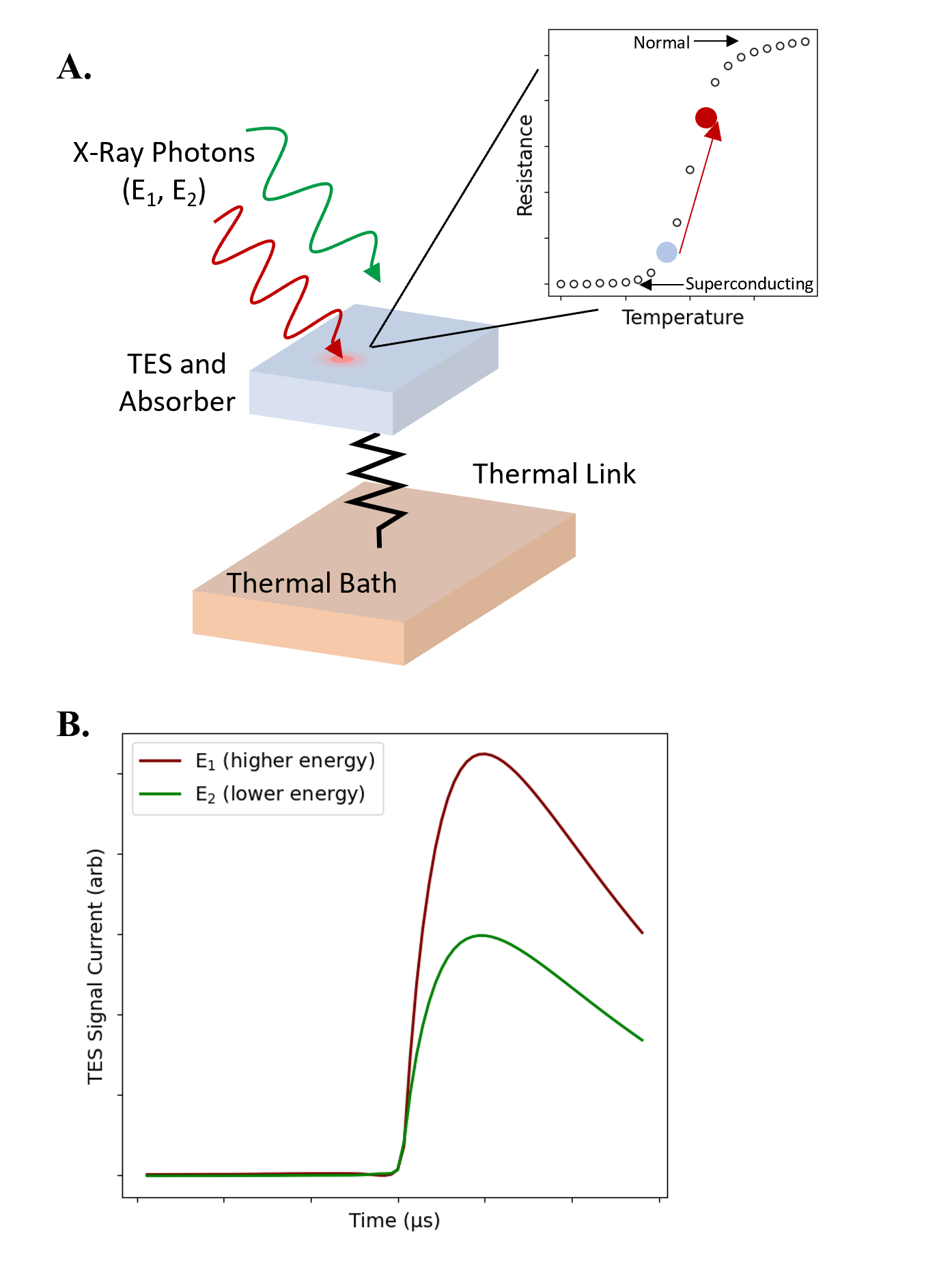}
\caption{(A)~Schematic of a TES. The TES and absorber are weakly thermally coupled to a silicon substrate serving as a thermal bath via a silicon nitride membrane. The TES is cooled into its superconducting state and voltage biased onto the superconducting-to-normal transition. When a photon is absorbed, the small increase in absorber and TES temperature results in a relatively large change in the TES resistance. Higher energy photons cause a larger change in temperature, and thus a larger change in the TES resistance. (B)~The change in TES resistance caused by photon absorption is read out as a negative-going pulse in the TES current (shown inverted here), with the pulse height proportional to the energy of the incident photon.}
\label{fig:tes}
\end{figure}

To isolate X-ray photons generated in the Pt target while still collecting enough photons for imaging, an X-ray detector with both high energy resolution and detection efficiency is required. TES microcalorimeters are an excellent choice for this application, as they bridge the gap between energy-dispersive detectors, which often have high efficiency but poor energy resolution, and wavelength-dispersive detectors, which have high energy resolution but poor efficiency.\supercite{Uhlig_TES_emiss_spec_2015} A TES microcalorimeter is a cryogenic superconducting detector which measures the energy of individual photons with excellent energy resolution.\supercite{ullom_review_2015} Each MINT TES consists of a 350~$\mu$m $\times$ 350~$\mu$m molybdenum/copper bilayer with eight noise-mitigating copper bars and a critical temperature (\emph{T$_c$}) of 130~mK.\supercite{Ullom_TESnoise_2004, pappas_tes2019, Doriese_TES_beamline_2017} A bismuth absorber 4.4~$\mu$m thick is deposited on top of each TES to improve X-ray absorption efficiency at higher energies. Each TES is suspended on a silicon nitride (SiN$_x$) membrane, which serves as a weak thermal link to a silicon substrate. The TES is cooled into a superconducting state, then voltage-biased into the superconducting-to-normal transition regime. When a photon is absorbed, it results in a small change in temperature of the coupled absorber-TES complex. Due to operating on the steep superconducting-to-normal transition, this small change in temperature results in a relatively large change in the measured TES resistance. The change in resistance from an absorbed photon is observed as a pulse in the measured current, with the pulse height proportional to the energy of the incident photon (Fig. \ref{fig:tes}). TES pixels are often combined into large, multi-pixel arrays to enable higher count rate measurements while maintaining high energy resolution. The MINT TES array consists of 240 pixels, and the pixels are readout using the time-division multiplexing (TDM) approach.\supercite{dekorte_timedivision_2003, Doriese_TDM_2016} The entire 240 pixel array is cooled to cryogenic temperatures to put the pixels into a superconducting state using a pulse-tube-backed two-stage adiabatic demagnetization refrigerator (ADR).\supercite{Hagmann_adr_1995, Doriese_TES_beamline_2017} Additional details on TES physics and operation can be found in a number of other publications.\supercite{ullom_review_2015, ullom_tesBeamline_2014, irwin_transitionedge_2005, Doriese_TES_beamline_2017}


\subsection{Source Term Monitor}
The source intensity can fluctuate throughout the measurement due to instabilities in the electron beam current or small variations in the Pt layer thickness at different dwell positions. It is critical to accurately measure this source intensity, as the detected counts must be normalized by the strength of the X-ray source to obtain accurate estimations of the attenuation through the sample. An EDS detector (Oxford Instruments) is located on the target-side side of the sample and operated continuously during data collection. The EDS detector monitors the total and Pt L$_\alpha$ count rates during each measurement and is used to quantify the X-ray source intensity.


\section{Data Processing} \label{sec:processing}

\subsection{TES Data Processing and Analysis}
\begin{figure}[!t]
\centering
\includegraphics[width=\linewidth]{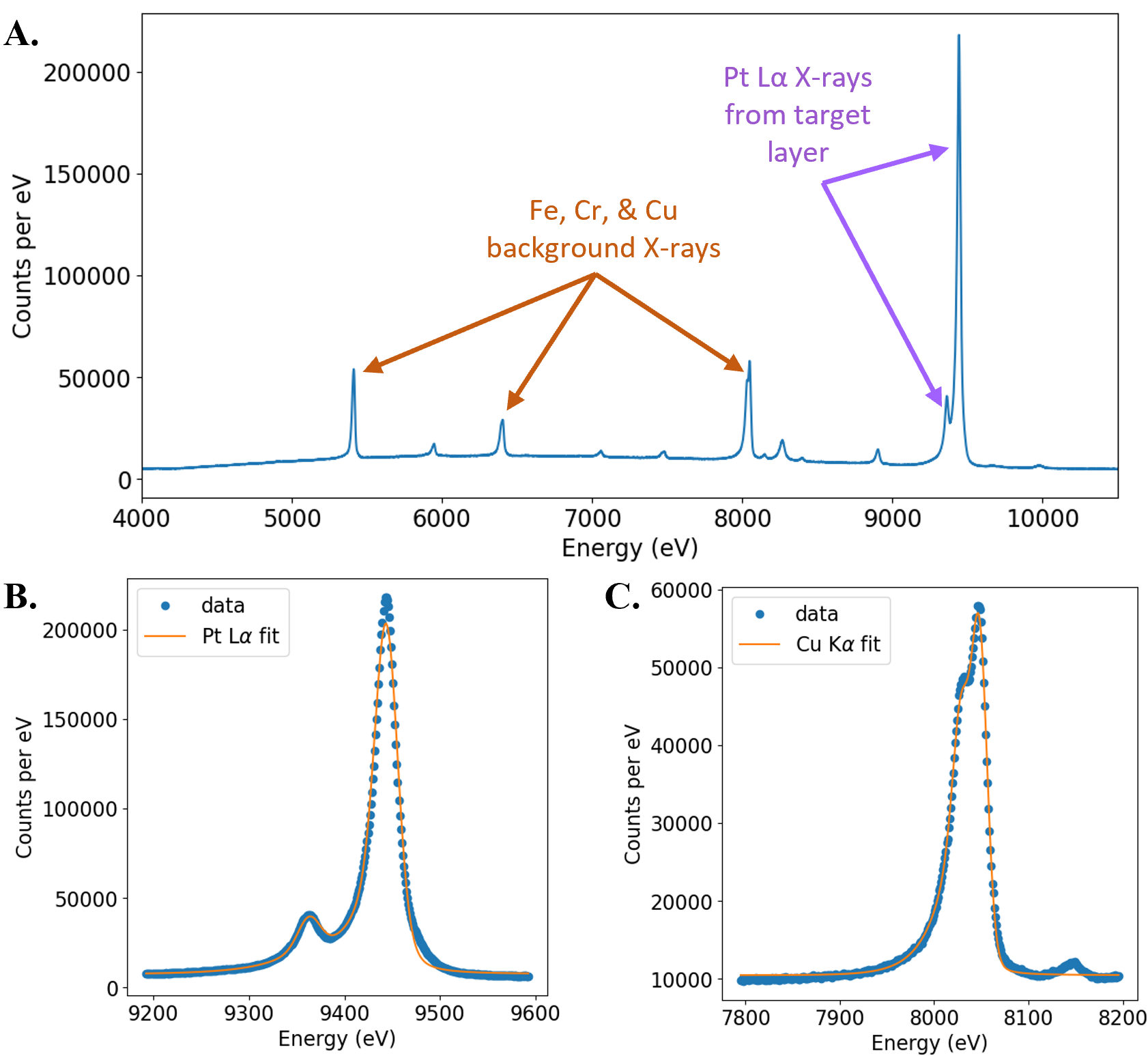}
\caption{
  (A)~Energy-calibrated TES spectrum combined over all TES pixels in the array and over all dwell positions. Purple arrows indicate the Pt L$_\alpha$ characteristic X-rays used for tomographic reconstruction. All other characteristic X-ray peaks are background peaks from sources such as the SEM chamber, sample holder, or cryostat. A selection of higher intensity background peaks are indicated by orange arrows. 
  (B)~Fit to the Pt L$_\alpha$ line for the full TES array over all dwell positions. This fit separates Pt X-rays generated in the target layer from bremsstrahlung background photons.
  (C)~Fit to the Cu K$_\alpha$ line for the full TES array over all dwell positions. The Cu K$_\alpha$ intrinsic line shape is well characterized,\supercite{Mendenhall2017,Hölzer_xraydata_1997} and this spectrum was used to establish the energy resolution of the TES spectrometer at 8 keV. 
}
\label{fig:TESdata}
\end{figure}

The broadband detection of the TES spectrometer allows for energies from approximately 4-12 keV to be collected simultaneously. The TES spectra can be used to isolate the Pt L$_\alpha$ fluorescence and select other energy bins of interest for use in tomographic reconstruction. Prior to the collection of CT data each day, a noise record of pulse-free TES current is obtained for each sensor. During CT data collection, each TES pixel in the array collects raw data pulses corresponding to X-ray events incident on the detector. Then, TES X-ray pulse and noise records are reduced to a time-lagged, energy-calibrated spectrum using the Microcalorimeter Analysis Software System (MASS).\supercite{fowler_practice_2016, Becker_analysis_2019} Additional details on the pulse processing steps can be found in the SM Section S3.A.1, as well as a number of previous publications.\supercite{fowler_practice_2016, Fowler2022, Szypryt_ebit_2019} The output of this processing is an energy-calibrated spectrum spanning the dynamic range of the TES spectrometer (Fig. \ref{fig:TESdata}A).

An X-ray spectrum is generated for each TES pixel at each dwell position during a tomographic scan, and information from the spectrum is input to a tomographic reconstruction code. Prior to use in reconstruction, each pixel is passed through a series of checks to ensure that lower quality data is excluded (SM Section S3.A.2). Extracting Pt L$_\alpha$ photons to use in the reconstruction can be done in two ways. First, the Pt L$_\alpha$ line can be fit to obtain the fluorescence counts at each dwell position for each TES (Fig. \ref{fig:TESdata}B). The fit consists of two Voigt functions of known line shape for the Pt L$_{\alpha_{1,2}}$ doublet\supercite{Zschornack_xraydata_2007} with exponential tails to account for the TES detector response function.\supercite{fowler_metrology_2017} This isolates only Pt L$_{\alpha_{1,2}}$ fluorescence and ensures that all photons used for reconstruction were generated in a nanoscale spot size. Another method to isolate photons generated in the target layer is to include all energy bins around the Pt L$_\alpha$ line, regardless of whether the photon included is from fluorescence or bremsstrahlung emission. This has the advantage of including more photons in the reconstruction, improving the signal-to-noise. However, it will include X-rays generated outside the Pt layer, degrading the spatial resolution. The amount of degradation depends on how many photons generated outside the Pt are included in the reconstruction and what their effective focal spot size is. We find that when fitting to the Pt L$_\alpha$ line, 600 million photons are included in the reconstruction. When binning from 9.1-10.1 keV and including the Pt L$_\alpha$ line, about 1 billion photons are included, increasing the photon counts by a factor of 1.67. In the 9.1-10.1 keV range, PENELOPE models indicate that 87\% of all detected photons originate in the Pt thin film. Therefore, a large proportion of the photons used still originate in the nanoscale X-ray spot generated in the Pt layer. In the reconstructions shown here, the coarser energy binning approach was used rather than the fit to the Pt L$_\alpha$ line. It is unlikely that this coarse binning approach will continue to be effective as the spatial resolution of MINT is pushed to the sub-100~nm level in future experiments, as the thinner targets required to generate smaller spots sizes will stop less electrons. Thus, we can expect a higher proportion of the bremsstrahlung to originate in materials other than the target and a larger negative impact on image quality.

The energy resolution of the coadded array was found to be 17.9~eV at the Cu K$_\alpha$ line (Fig. \ref{fig:TESdata}C). The Cu K$_\alpha$ line was used to measure resolution because its line shape is well known,\supercite{Mendenhall2017,Hölzer_xraydata_1997} This energy resolution is sufficient to separate the Pt L$_\alpha$ counts from the background with high signal-to-noise, but does not represent the state-of-the-art for TES pixels. TES pixels are capable of achieving much higher energy resolution than what is demonstrated here, with resolving powers ($E/\Delta E$) greater than $1000$.\supercite{ullom_review_2015} The current TES array was optimized for faster photon detection to be compatible with the X-ray generation rate of the SEM, at the cost of a minor degradation in the energy resolution.\supercite{pappas_tes2019}


\begin{figure*}[!t]
\centering
\includegraphics[width=\textwidth, keepaspectratio]{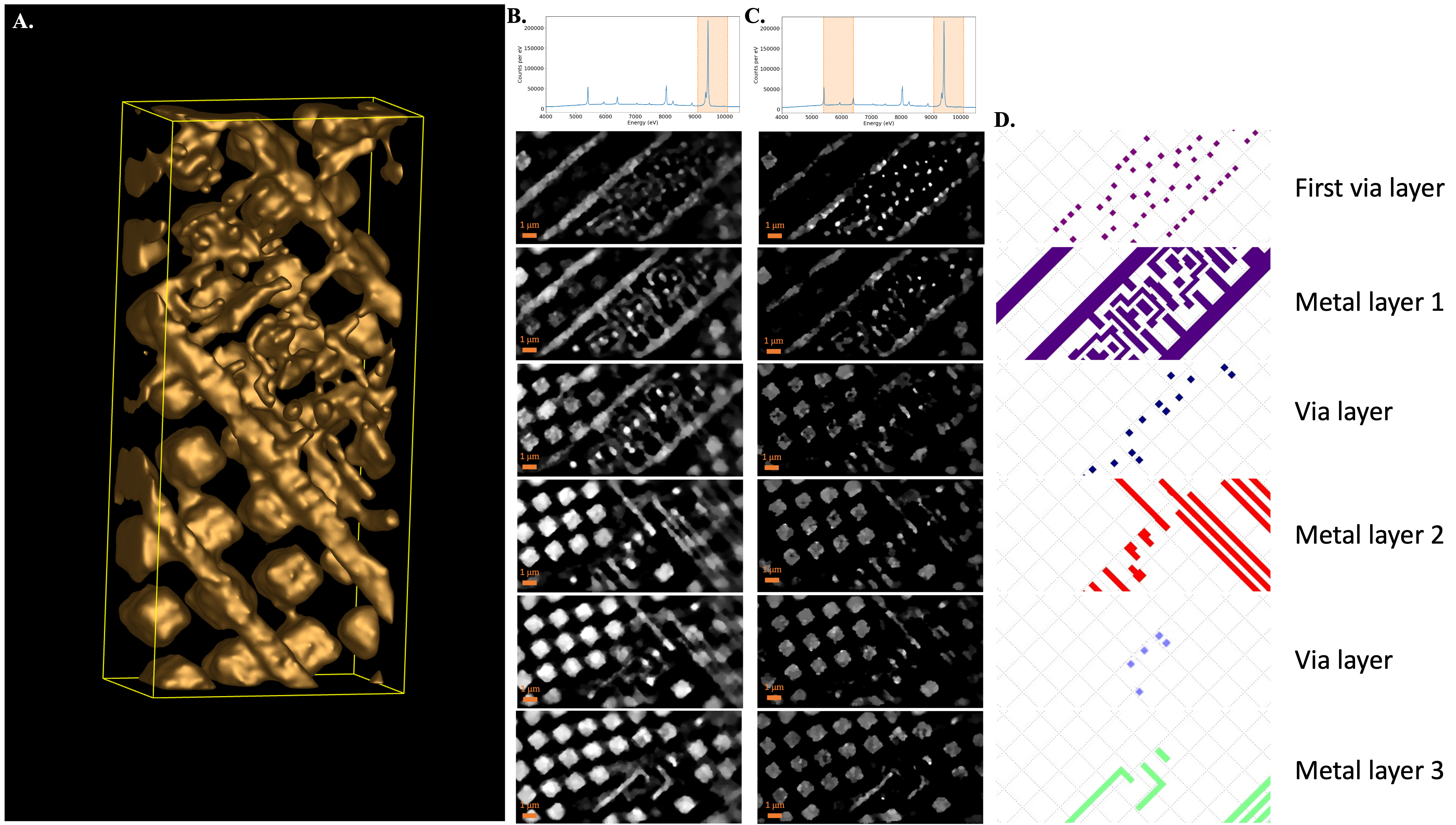}
\caption{(A) three-dimensional reconstruction of an IC fabricated at the 130~nm node, using X-rays in the 9.1-10.1~keV energy band. This band includes all Pt L$_\alpha$ photons. (B) Spectrum from the TES detector with the 9.1-10.1~keV energy band used for reconstruction highlighted in orange (top). Reconstruction results, separated by IC layer, are shown under the spectrum. These slices were taken from the reconstruction shown in A. (C) Multi-energy reconstruction results, using the 9.1-10.1~keV and the 5.4-6.4~keV band, shown under the TES spectrum with the X-ray energies used highlighted in orange. Here, only the first via layer is resolved more clearly than when only using 9.1-10.1~keV photons, indicating a material other than Cu may be present. (D) GDS ground truth for each of the metal via and wiring layers, for comparison with reconstruction results. A portion of this figure has appeared in Ref.~\cite{levine2023tabletop}}
\label{fig:tomo}
\end{figure*}

\subsection{Image Generation}
Both two-dimensional radiographs and three-dimensional tomographic reconstructions were generated from the collected data. Details on the data collection can be found in the SM, Section S4. Due to a combination of positional drift and offset in the commanded and executed angular and linear positions of the stage, radiographs collected during each scan can be offset. The offset was found to be minor between inner and outer regions collected at the same angle, but were significant across different angular projections. If not accounted for, this misalignment would degrade image resolution in the three-dimensional reconstruction. A radiograph-alignment software was developed to quantify and correct for these positional offsets. The software operates by first aligning the two-dimensional radiographs from a scan at a given projection angle. Then, angle-to-angle alignments are performed in a sequential fashion with adjacent angles aligned to one another (e.g., $0^\circ$ to $\pm7.5^\circ$, $\pm7.5^\circ$ to $\pm15^\circ$, etc.).

To generate a three-dimensional image from aligned two-dimensional radiographs, the code TomoScatt was used.\supercite{Levine_tomoscatt_2019} TomoScatt utilizes an objective function with a Bayesian prior and that evaluates the log likelihood that a given reconstruction is optimal.\supercite{SauerBouman_1993} The code was edited here to accommodate a finite source and aspects of the TES spectrometer, such as pixel positioning and energy-specific reconstructions. Additional information regarding reconstruction algorithm development and results can be found in a recent publication.\supercite{levine2023tabletop}

\section{Reconstruction Results} \label{sec:results}

The sample IC contains three metal wiring and three metal via layers with minimum feature sizes of 160~nm. Two reconstructions are shown, one using the 9.1-10.1~keV energy band including the Pt L$_\alpha$ line (Fig. \ref{fig:tomo}A-B) and one using both the 9.1-10.1~keV band and the 5.4-6.4~keV band (Fig. \ref{fig:tomo}C). These reconstructions are compared to the wiring layers shown in the ground truth graphic design system (GDS) file (Fig. \ref{fig:tomo}D), but the GDS file was not used as an input to either reconstruction. The larger cubes in the reconstructions are chemical-mechanical polishing (CMP) filling. CMP fill is included in ICs during the fabrication process to provide thermal and mechanical stability\supercite{Zhao_cmp_2013}; it is not included in the GDS file.

Multiple reconstructions are shown to demonstrate the spatial resolution of MINT as well as first steps towards element-sensitive reconstructions. The expected contrast for a given material will change based on what energy of photons are used in the reconstruction (SM, Section S6). For Cu features it is advantageous to use X-ray energies just above the Cu K-edge, making 9.1-10.1~keV an ideal choice. For other common IC materials, such as Ta or W, using lower energy bands will be advantageous for improving contrast. However, the fraction of X-rays originating in the nanoscale spot in the Pt target layer decreases as lower energy bands are included. For the 5.4-6.4~keV band, only 51\% of the detected photons are generated in the Pt target. Additionally, lower energy bands offer less contrast for Cu imaging than the 9.1-10.1~keV band. So, while including the lower energy band may improve contrast for non-Cu materials, it will decrease contrast for Cu and lead to an increase in the X-ray focal spot size.

The first reconstruction, using only the 9.1-10.1~keV band, resolves all expected features in the IC. Wiring and via layers are clearly distinguished from the surrounding CMP fill. This is expected, as the majority of photons included in the reconstruction are generated in the Pt target and are at energies which provide a high Cu-SiO$_2$ contrast. This demonstrates that MINT can resolve features at least as small as 160~nm. The second reconstruction, which adds the 5.4-6.4~keV energy band into the reconstruction, is less effective at imaging most layers. However, it is more effective at resolving the first via layer, demonstrating higher contrast of the vias than the reconstruction using only the higher energy band. This suggests that the first layer is likely not comprised of Cu, and is potentially W or Ta. Both W and Ta fluorescence peaks are visible in the TES spectrum, and both may be present in the IC. While future work is needed to fully explore spectral imaging capabilities in MINT, the TES spectrometer offers several advantages for this imaging mode moving forward. The high resolving power of the TES allows many energy bands to be selected, offering the capability to identify many different materials simultaneously. Additionally, data across a broadband spectral range is collected in the same scan, so multiple reconstructions including different energy bins can be made from a single tomography dataset. These results demonstrate the capability of MINT to achieve nanoscale spatial resolution and elemental sensitivity in a compact instrument, and indicate the promise of utilizing SEM and TES-based systems for X-ray nanotomography.


\section{Future Outlook} \label{sec:future}
The current implementation of MINT is a proof-of-concept, demonstrating the promise of an SEM and TES-based approach to nanoscale tomography. The first generation of the tool as shown here achieves spatial resolution comparable to state-of-the-art laboratory X-ray CT instruments, with additional capabilities for spectral imaging. MINT also has a clear path forward to further improve spatial resolution, imaging speed, and spectral imaging capabilities. These upgrades involve improvements to the hardware, target design, and reconstruction approach, but do not alter the overall MINT concept.

First, the electron column could be upgraded to achieve higher beam currents at smaller spot sizes, improving both the imaging speed and resolution. Modern electron columns can achieve microampere beam currents, significantly higher than what was used for the demonstration presented here.\supercite{Krysztof_eGun, Kusunoki_eGun}. Additionally, at the 160~nm spatial resolution goal, the target can be designed such that the majority of X-rays are generated in the target layer. In this limit, a commercial megapixel camera could be used to drastically improve imaging speed, at the cost of elemental-sensitive reconstructions. Using an example commercial X-ray detector (Dectris Eiger2), we estimate that the imaging speed could be approximately 850 times faster than the current 240-pixel TES spectrometer. However, at spatial resolution goals below 100~nm, it will become difficult to maintain a thin enough target layer, small enough electron spot size, and sufficient electron beam current for imaging while keeping the majority of X-ray generation contained to the thin film metal target. This will result in slow imaging speeds, or a degradation in the achievable spatial resolution as the beam current is increased. One approach to push the spatial resolution to the level of feature sizes in state-of-the-art ICs is to isolate photons from one or multiple nanosized X-ray targets, which would contain the fluorescence X-ray emission from each material to a nanoscale spot size regardless of the size and current of the incident electron beam. Here, an energy resolving detector would be essential to isolating the fluorescence from each nanotarget. A TES spectrometer could then be used to isolate X-ray emission from specific fluorescence lines of interest from mutliple nanofeatured materials.\supercite{lavely_nanopatterns2022}

In addition to source upgrades to improve imaging speed and resolution, advances in TES detector technology will continue to improve the energy resolution while drastically improving the collection efficiency, enabling faster imaging speeds while maintaining elemental selectivity. These advances include individual TES pixels with higher count rate capabilities and higher pixel count TES spectrometers which collect over a larger solid angle. Already, advances in TES multiplexing technology have enabled the development of spectrometers with upwards of 1000 pixels,\supercite{Szypryt_1k_2023} with a 3000-pixel spectrometer in development to dramatically increase photon collection capabilities in MINT.\supercite{Szypryt_3k_2021} If integrated in the current tool, it would immediately improve the imaging speed by a factor of 14. Lastly, the use of an energy resolving detector creates the opportunity for spectral imaging approaches that could determine the elemental or chemical composition of the sample. A proof-of-concept for element-specific reconstructions with a TES was demonstrated here, but more advanced spectral tomography cases at synchrotron beamlines have resulted in the combination of tomographic and spectroscopic techniques to yield information on chemical states within each voxel.\supercite{feng_spectralCT_2020, howard_spectralCT_2020, DeSamber_spectralCT_2008} TES spectrometers are widely used for X-ray spectroscopy\supercite{ullom_review_2015, Uhlig_TES_emiss_spec_2015} and have a history of bringing synchrotron techniques to the laboratory bench,\supercite{Miaja-Avila_TES_laser_2016} creating the opportunity to develop advanced tabletop spectral tomography approaches centered around the MINT concept.

\section{Conclusion}
MINT represents an innovative approach to X-ray tomography, changing many of the limitations on the X-ray source, system geometry, and X-ray detection than hinder conventional laboratory CT instruments. MINT combines a scanning electron microscope (SEM) with a transition-edge sensor (TES) spectrometer to enable X-ray tomographic imaging with nanoscale spatial resolution and elemental specificity. The tightly focused electron beam spot generates X-rays in a metal target layer, maintaining a small focal spot size and eliminating the need for X-ray focusing optics. The broadband detection and energy resolution of the TES array is then used to select for specific energy bands to emphasize spatial resolution or detection of different materials in the sample. A proof-of-concept measurement indicates that MINT can resolve features down to 160~nm in a planar Cu-SiO$_2$ integrated circuit (IC) sample with a limited number of angular projections, with great potential for future improvement. MINT is inherently flexible, with the ability to adjust X-ray target materials, electron beam parameters, and geometric magnification to achieve optimal imaging conditions for a given spatial resolution goal or sample composition. The current implementation of MINT demonstrates the promise of merging an SEM with energy-resolved X-ray detection for nanoscale tomography, the next generation could achieve nanoscale, element-specific tomography not previously achieved in laboratory X-ray CT instruments.

\section{Acknowledgements}
The authors would like to thank Eugene Lavely, Adam Marcinuk, Paul Moffitt, Steve O'Neil, Thomas Stark, Chris Willis, and others at BAE Systems for their role in the MINT concept development and initial instrument integration in the NIST Boulder Laboratories. The authors would also like to thank Damien Griveau, Julien Alberto, Jeremie Silvent, Elodie Verzeroli, and others at Orsay Physics for their support in SEM maintenance and operation. Sandia National Laboratories is a multimission laboratory managed and operated by National Technology and Engineering Solutions of Sandia, LLC, a wholly owned subsidiary of Honeywell International Inc., for the U.S. Department of Energy's National Nuclear Security Administration under contract DE-NA0003525. The information, data, or work presented herein was funded in part by the Office of the Director of National Intelligence (ODNI), Intelligence Advanced Research Projects Activity (IARPA), via agreements D2019-1908080004, D2019-1906200003, D2021-2106170004, and FA8702-15-D-0001. NN and PS were supported in part by National Research Council Postdoctoral Fellowships. This work was performed in part under the following financial assistance award 70NANB18H006 from the U.S. Department of Commerce, National Institute of Standards and Technology. This article describes objective technical results and analysis. Mention of commercial products does not imply endorsement by the authors or their institutions. The views and opinions of authors expressed herein do not necessarily state or reflect those of the United States Government or any agency thereof.

\section{Contributions}
Hardware development: JZH, DM, NJN, NJO, CGP, DSS, PS, JWW.
Software development: JWF, JZH, DM, GCO, CW, JWW.
TES array development: DAB, WBD, GCH, KMM, GCO, CGP, DRS, DSS, PS, LV, JCW.
TES readout and cryostat development: DAB, WBD, MD, JWF, JDG, GCO, CDR, PS, JNU.
Data collection: ALD, JZH, NJN, PS.
Data processing and analysis: JWF, JZH, ZHL, NJN, PS, KRT, CW.
Radiation transport simulations: NJN.
Sample and target development: ALD, NJN.
Tomographic reconstructions: BKA, JWF, DTF, RNG, ESJ, BLK, ZHL, PAS, KRT, CTV, CW.
Writing: NJN.
Figure preparation: NJN.
Management: ESJ, KWL, DSS, JNU.

\section*{Conflict of Interest}
ZHL declares Z. H. Levine, ``Efficient Method for Tomographic Reconstruction in the Presence of Fresnel Diffraction,'' U.~S.~Patent Application 17/700,884. 
All other authors declare no competing interests.

\printbibliography
\end{document}


\maketitle
\section{Imaging Speed}
A derivation of Equation 1 is given here, describing how the relative imaging speed was determined. The $N$ measured photons consist of $B$ background bremsstrahlung photons plus $F$ fluorescence photons in the peak, with
\begin{equation}
    F = N - B.
\end{equation}
As $F$ is the difference between two independent Poisson random deviates, it follows the Skellam distribution~\supercite{Skellam1946}, with a variance
\begin{equation}
    \sigma^2(F) = {E[N] + E[B]} \approx F+2B.
\end{equation}
Where E[] represents the expectation value of the enclosed variable. We define $f$ to be the average count rate of fluorescence photons, $b$ to be the average bremsstrahlung count rate per eV, and $t$ to be the integration time. Then, $E[F] = ft$, and the uncertainty on our estimation of $F$ can be expressed as
\begin{equation} \label{eq:shotNoise}
    \sigma(F) = \sqrt{ft+2b\Delta E t}
\end{equation}
where $\Delta E$ is the signal bandwidth in eV---a combination of the intrinsic emission-line width and the resolution of the TES spectrometer at the fluorescence line of interest.

The goal of transmission-based X-ray tomography is to distinguish different materials in each voxel by their X-ray transmission. The transmission is $T = \exp(-\mu x)$ through a voxel of thickness $x$ given an X-ray absorption of $\mu$ per unit length. Considering only one voxel, if $F_0$ were the expected number of fluorescence photons that would be seen if that voxel had zero absorption ($\mu=0$) and all other layers of voxels were unchanged, the expected value of $F$ would be
\begin{equation}
    E[F] = F_0\exp(-\mu x).
\end{equation}
To first order, the relationship between uncertainties on $F$ and $\mu$ would be
\begin{equation}
    \sigma(F) = xF\sigma(\mu).
\end{equation}
We define $\Delta\mu$ as the intrinsic contrast between Cu and SiO$_2$. Thus, a measurement with $\sigma(\mu) < \Delta\mu/k$ will have a signal-to-noise ratio of $k$ for the Cu/SiO$_2$ discrimination task. The requirement is that the absorption contrast meets the condition
\begin{align}
    \Delta\mu &> k\sigma(\mu) = \frac{k\sigma(F)}{xF} \\
    & > \frac{k}{x}\frac{\sqrt{ft+2b\Delta E t}}{ft} \\
    & > \frac{k}{x}\sqrt{\frac{1+2b\Delta E/f}{ft}}.
\end{align}
Rewriting in terms of imaging speed $1/t$, we get
\begin{equation}
    \frac{1}{t} < f \left( \frac{x\Delta\mu}{k}\right)^2 (1+2b\Delta E/f)^{-1}.
\end{equation}

The values of $x$ and $k$ are not relevant to the calculation of relative imaging speed given in the main text, and thus are omitted. When predicting the absolute imaging speed, the speed will scale as the square of the voxel thickness. Likewise, the time required grows as the square of the required per-voxel signal-to-noise.


\section{Sample Preparation} \label{sec:samplePrep}
\begin{figure}[!b]
\centering
\includegraphics[scale=0.7]{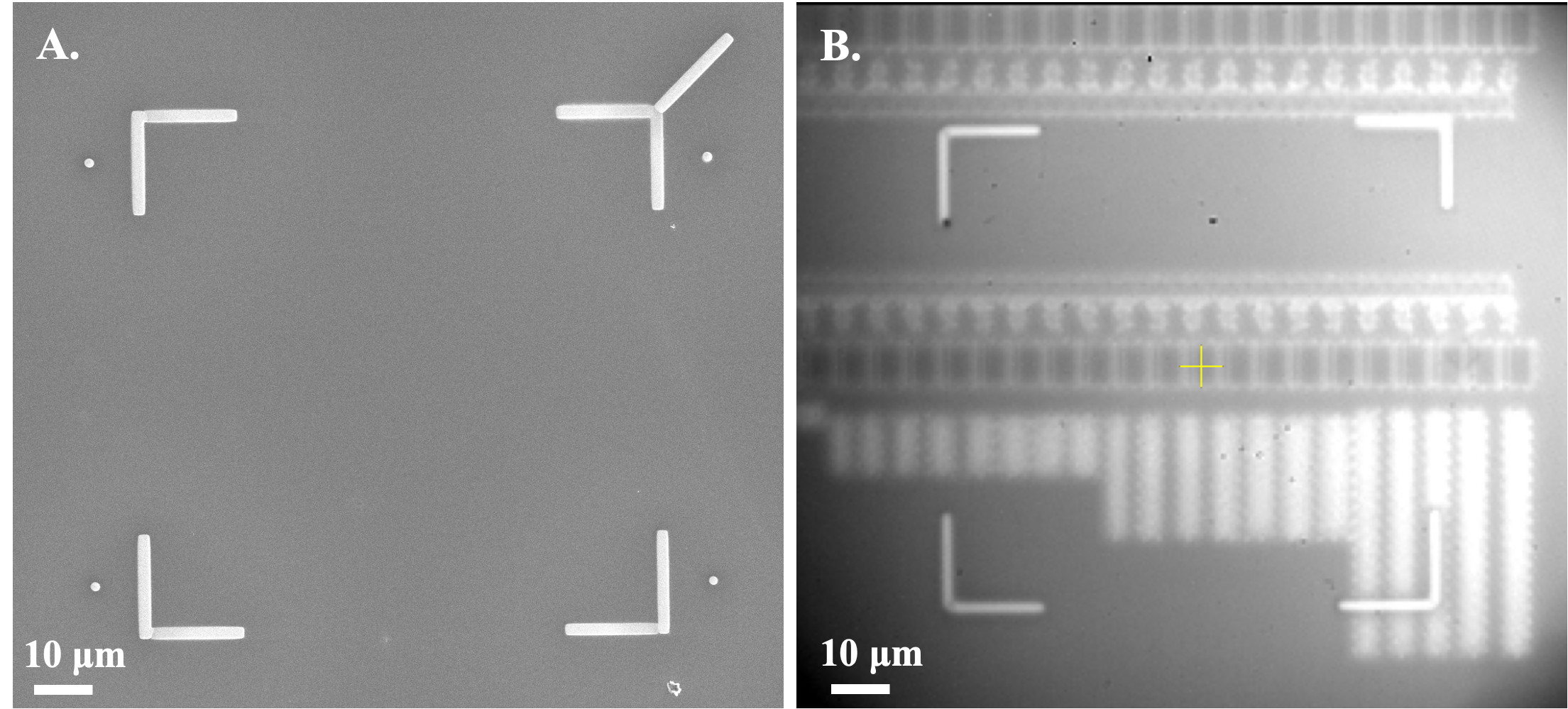}
\caption{(A)~SEM image of the Pt target and W fiducials (B)~Infrared (IR) image of the sample surface, showing the W fiducial position relative to IC features. The tomographic region of interest is within the area enclosed by the W fiducials.}
\label{fig:fiducials}
\end{figure}

Preparation of the demonstration IC sample was required to achieve the desired system geometry for high magnification. The sample consists of Cu wiring in an SiO$_2$ dielectric on a Si substrate. First, the Cu/SiO$_2$ surface was polished (Allied Multiprep polisher) and surface topography was measured (Dektak profilometer, Bruker). Then, metal layers of the IC were removed via spin milling in an FEI G4 plasma focused ion beam (FIB) system. The metal layers removed from the IC contained larger Cu features and were not of interest for a nanotomography demonstration. Removing these larger wiring layers is not strictly necessary, but it enables faster imaging of the remaining layers. After layer removal, three metal wiring and three metal via layers remain. The Cu/SiO$_2$ surface was then mounted on a 60 $\mu$m thick glassy carbon layer (SPI Supplies) using Epotek 377 epoxy to provide structural rigidity.

The Si substrate on the backside of the IC was thinned to the desired thickness, in this case 8.5~$\mu$m. The thickness of the Si layer, referred to as the spacer layer, depends on the intended spatial resolution of the measurement (see Section II.B). To thin the spacer, the surface was lapped (Allied Multiprep) and polished, with the thickness verified using an Allied Vision System reflectometer. The glassy carbon-facing side of the sample is then bonded to a graphite puck using Epotek 377. The graphite puck serves as a sample carrier for further processing and insertion into the MINT sample holder. The puck has a window cut into the back aligned with the sample mounting position to reduce photon attenuation.

Once on the graphite puck, fiducial marks of tungsten are deposited on the Si surface using the FEI G4 plasma FIB (Fig. \ref{fig:fiducials}A). The fiducials are deposited around the scan region of interest (ROI) and are used both for macroscale localization of the ROI and for positional corrections during tomographic data collections (Fig. \ref{fig:fiducials}B). The X-ray target is then deposited directly over these fiducials and onto the spacer layer. For the current IC sample, 10 nm of Cr was evaporated as an adhesion layer followed by 100~nm of Pt.


\section{Data Processing}
\subsection{TES Data Processing}
\subsubsection{Pulse Processing}
The steps for TES data processing are described in more detail here. First, a series of cuts are performed to remove anomalous pulses. The majority of pulses cut are excluded due to pulse pileup effects, in which a second pulse arrives during the thermal decay of a previous pulse. After irregular pulses are removed, an average pulse is created for each pixel and used along with the noise autocorrelation function to create an optimal filter that maximizes the signal-to-noise ratio (SNR) during pulse height estimation.\supercite{szymkowiak_filter_1993, anderson_filter_1979}

The measured pulse height in each pixel can be affected by a number of other physical effects beyond the energy of the incident photon, namely temperature drift on the (unregulated) 1 K stage of the ADR and the arrival time of a photon relative to the digital sampling clock. We correct for temperature-dependent drift with an algorithm that sharpens the energy spectrum by minimizing spectral entropy. It uses the mean pretrigger period of each pulse record as an indicator of temperature drift; sensor gain is found to vary linearly with this drift. Various approaches can be take to mitigate the arrival-time bias; most importantly, we smooth the optimal filter by a 1-pole filter with a 3 dB point of approximately 5\,kHz to reduce arrival-time sensitivity. Additional details on these biases and corrections can be found in a prior publication.\supercite{fowler_practice_2016}

The optimally filtered, temperature drift-corrected, and arrival time-corrected pulse heights are  calibrated into energy using known fluorescence lines. Here, the Pt L$_\alpha$, Pt L$_\beta$, Cu K$_\alpha$, and Cr K$_\alpha$ lines are used as calibration anchor points. Each line is fit to a model consisting of a known line shape\supercite{Zschornack_xraydata_2007, Hölzer_xraydata_1997} and a cubic spline is used to create a gain function relating the pulse height to X-ray energy.\supercite{Fowler2022} The end result of these processing steps is a series of time-lagged, energy-calibrated X-ray pulses, which can be used to generate an energy-calibrated spectrum.

\subsubsection{TES Pixel Screening}
Prior to including data in the tomographic reconstruction, each pixel is screened to ensure that it is operating as intended. Each TES is screened based on the number of cut pulses, the Pt L$_\alpha$ count rate, and the energy resolution at the Pt L$_\alpha$ line. A TES pixel is flagged and excluded from the dataset if more than 20\,\% of detected pulses were cut, if the Pt L$_\alpha$ count rate is more than 25\,\% higher or lower than the median rate across all pixels, or if the energy resolution at the Pt L$_\alpha$ line is worse than 60~eV (FWHM). This eliminates pixels with manufacturing defects or errors in TES configuration that cause irregular pulse shapes or poor energy resolution. On average, 193 of the TES pixels were used in a given scan. In the TES pixels that passed these checks, 94\,\% of X-ray pulses were kept and used for subsequent analysis steps.


\begin{figure}[!t]
\centering
\includegraphics[scale=0.6]{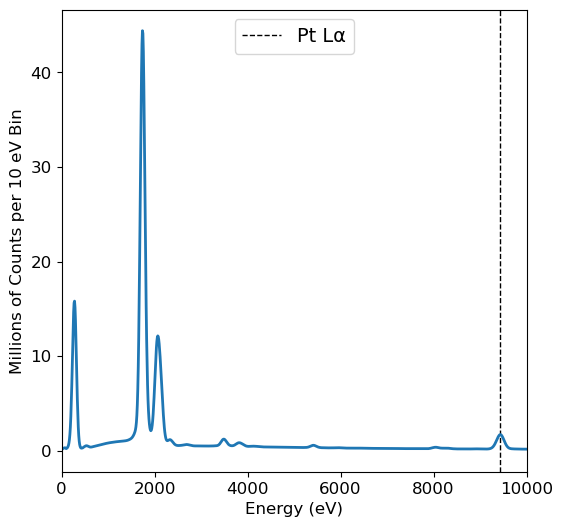}
\caption{Spectrum collected by the EDS detector summed over all dwell positions. The Pt L$_\alpha$ line (vertical black dashed line) is fit and used to estimate fluctuations in the X-ray source during data collection. These fluctuations can arise from small variations in the Pt layer thickness over the scan region or fluctuations in the electron beam current.}
\label{fig:eds}
\end{figure}

\subsection{EDS Data Processing and Analysis}
The energy dispersive spectroscopy (EDS) detector is used as a monitor of fluctuations in the X-ray source term to normalize the detected transmission signal from the TES spectrometer. The X-ray source intensity can vary both spatially and temporally due to variations in the target thickness and SEM beam current. During MINT data collection, the EDS data is saved every second and consists of the detected counts per 10 eV bin, the timestamps at the beginning and end of each collection, and the detector's effective live time. Due to the high count rates incident on the EDS detector, pulse pileup occurs which renders certain subsets of the EDS data unusable by the Aztec Software system. The effective live time is the portion of the dwell not affected by pulse pileup. Here, the typical EDS live time was 0.75 seconds per 1 second dwell. The EDS timestamps are used to align the EDS spectra with the corresponding tomographic dwell and its TES and positional data. Then, all EDS spectra contained in a given dwell are summed to generate a final spectrum (Fig. \ref{fig:eds}). The final EDS spectrum for a given dwell has an energy resolution of approximately 200~eV at the Pt L$_\alpha$ line and tracks the source intensity at each tomographic dwell position. The Pt L$_\alpha$ count rate was used for X-ray source term normalization during tomographic reconstructions, as it encompasses both fluctuations in the SEM beam current and variations in the target thickness at each dwell position. The Pt L$_\alpha$ count rates were found by fitting to the fluorescence line for each dwell and dividing by the live-time corrected dwell time. 


\subsection{Positional Data}
Achieving nanoscale spatial resolution requires similarly precise knowledge of the source position on the sample. During each tomographic scan, the $x$, $y$, and $z$ stage positions are stored relative to the IC sample plane-of-reference for both the beginning and end of the dwell period. This provides known electron beam positions for the start and end of the dwell, allowing the magnitude of the drift during each dwell period to be easily calculated. Additionally, the rotation angle of the stage is saved at each dwell.


\subsection{Data Consolidation}
MINT requires the collection of data from multiple sources, namely the stage, SEM, TES spectrometer, and EDS detector. To facilitate their use in tomographic reconstruction codes, it is advantageous to consolidate the data into one input file. MINT data products relevant to tomography were stored in a Hierarchical Data Format version 5 (HDF5) file, with one HDF5 file per inner or outer scan. These HDF5 files include the TES Pt L$_\alpha$ and total count rates per pixel per dwell, the EDS total and Pt L$_\alpha$ count rates per dwell, the stage positions at the beginning and end of each dwell, the TES pixel positions relative to the center of the TES array, and timestamps for the beginning and end of each dwell, along with various other TES diagnostics not used for tomographic reconstructions. In total, the tomographic reconstruction codes utilized 74 HDF5 files over the angular range with 18 outer region scans and 56 inner region scans.

\section{Scan Strategy and Implementation}
\begin{figure}[!b]
\centering
\includegraphics[width=8cm]{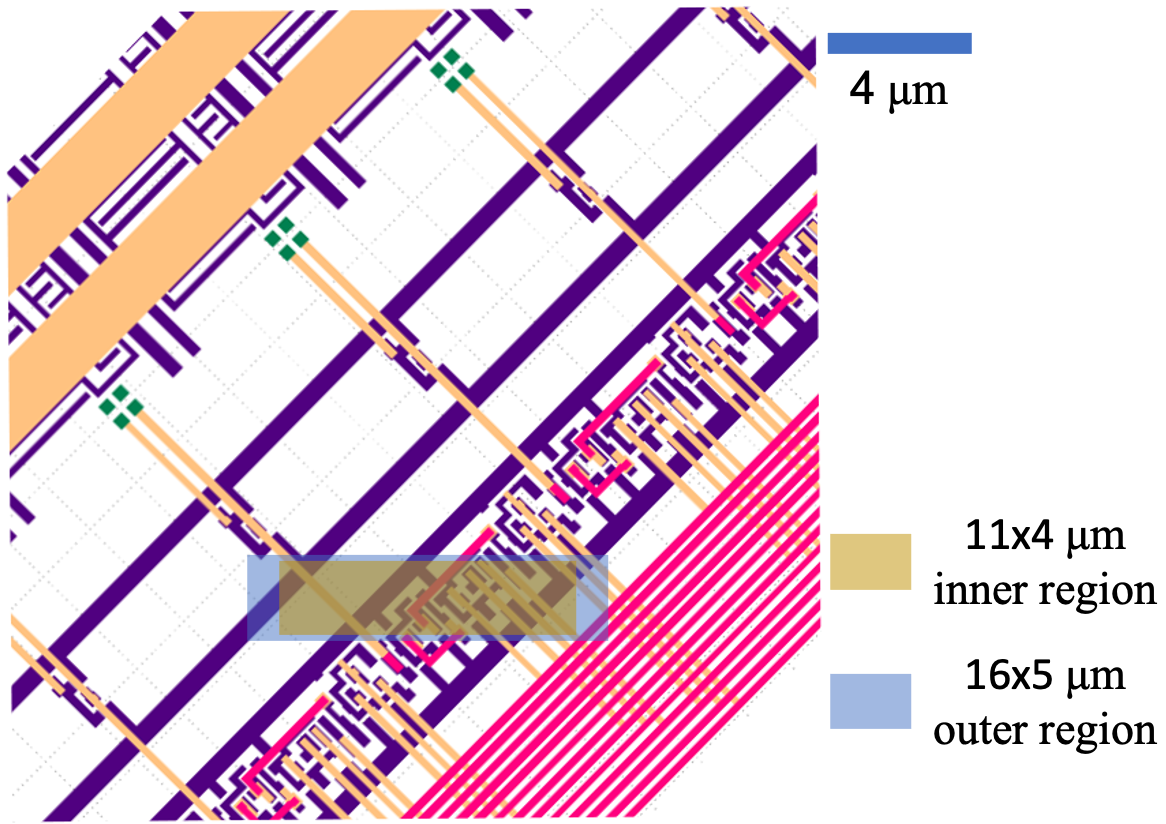}
\caption{Metal layers of the demonstration IC sample. The sample was thinned to contain three metal (yellow, purple, pink) and three via layers (not shown). The green features extend through multiple metal and via layers. The inner (yellow) and outer (blue) rectangular scan regions were chosen to cover the digital logic region of the IC, which contains the densest population of 160~nm Cu features.}
\label{fig:gds}
\end{figure}

A scan plan for tomographic data collection was developed based on MINT system parameters. The ADR hold time at 100~mK limited the data acquisition time per day to approximately 18 hours. The scan area, step sizes, and dwell time at each step were chosen based on this time allotment. To maximize the amount of time spent collecting data during the ADR hold time, one angular projection was collected each day and setup for a new angular projection was performed during the ADR magnet cycle, which takes approximately four hours to complete. Data at each angle was collected over a series of scans of a smaller inner region and a larger outer region, with the feature-dense digital logic region approximately centered in each (Fig.~\ref{fig:gds}). Three 11~$\mu$m $\times$ 4~$\mu$m inner region scans were collected to provide high spatial resolution in the digital logic region. In the inner region scans, a step size of 405~nm was used, with a dwell time of 45 seconds per dwell. The exact dwell positions between each inner region were offset slightly to provide a denser coverage of the region-of-interest (ROI) across all three inner region scans. After the inner region scans were complete, one 16~$\mu$m $\times$ 5~$\mu$m outer region scan was collected at a step size of 700~nm and dwell time of 65 seconds. The outer region is collected to mitigate edge effects in the inner region scans, which degrade the spatial resolution on the outer edges of the ROI.\supercite{Kyrieleis_ROItomo_2011, Ravishankar_ROItomo_2004} As the sample rotates to higher angles around the $y$-axis, the projection of the ROI extends in the $x$-axis. This leads to a larger ROI at higher angles. To complete all four scans within the ADR hold time and maintain the same dwell density within the rotated scan regions, the step sizes and dwell times were adjusted from the values given above.

During a scan, corrections must be made periodically to account for positional drift of the electron beam or stage to ensure that the data is collected at the desired dwell location. To accomplish this, SEM images of a W fiducial at the same commanded stage location are taken before and after the first dwell position and after each subsequent dwell. Each post-dwell image is compared and aligned to the pre-dwell image, with the amount of alignment needed corresponding to the amount of positional drift which occurred during that dwell. Any positional offsets are integrated into the next dwell position to correct for drift. This process of fiducial alignment and positional corrections adds approximately 20 seconds of overhead per dwell. Details on the sample positioning system and positional drift during scanning can be found in Sections S5-S6.

Data were collected over a total of 12 angles, ranging from $-37.5\degree$ to $+45\degree$ in $7.5\degree$ angular steps. The angular step is set by the angular coverage provided by the x-ray detector. In MINT, a spectrum is collected at each dwell position in the scan region. A reconstruction using the collected spectra at a single projection angle is equivalent to $7.5\degree$ of angular coverage. The total angular range is set by limitations in the sample and system geometry. In general, the angle of rotation is limited for planar samples due to the increasing path length to exit the sample at higher rotation angles. This is a challenge facing CT scans on planar samples in all tomography systems, as the increased attenuation at higher path lengths through the sample can drastically reduce the number of transmitted photons available for image generation. In MINT, two additional challenges for planar samples arise due to the unique system configuration. First, a risk of stage collision with the SEM column exists at angles larger than $\pm45\degree$. Second, at angles more negative than $-37.5\degree$ the sample holder obscures the EDS detector and the source term cannot be estimated. The angular step size was selected based on the solid angle of the TES array, which provides $7.5\degree$ of angular coverage.


\section{Stage and Sample Holder}

\begin{figure}[!b]
\centering
\includegraphics[scale=0.8]{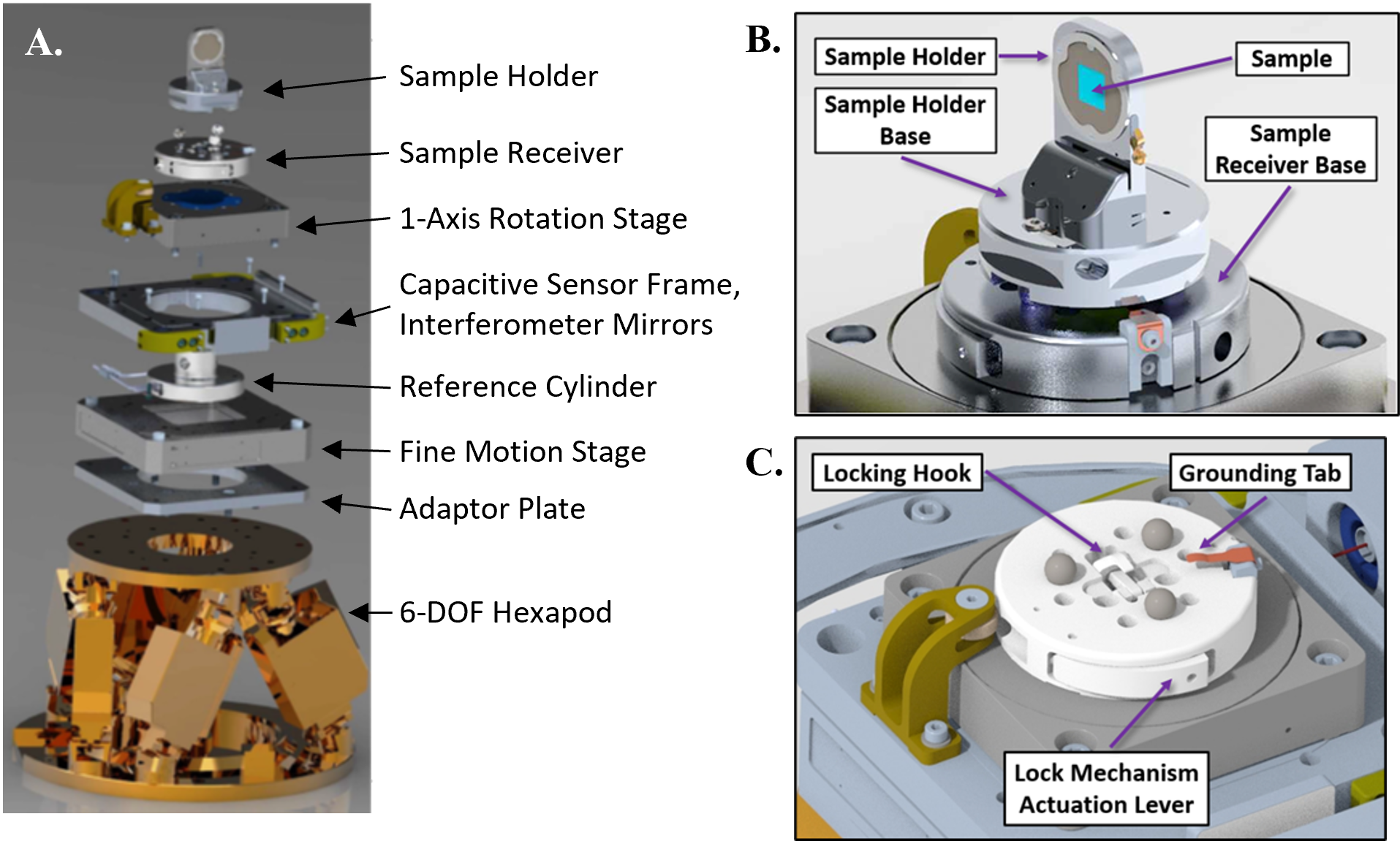}
\caption{(A) The sample positioning stack, consisting of a three-stage sample positioning system (SPS), external capacitive and interferometric sensors, and custom-made sample holder and receiver. (B) The sample holder assembly and sample, attached to the sample receiver base. (C) The sample receiver base, which locks the sample holder securely to the SPS during tomographic data collection.}
\label{fig:stage}
\end{figure}

To perform tomography, a sample positioning system (SPS) which enables high-precision \textit{x}, \textit{y}, \textit{z}, and angular motion is required. The MINT SPS consists of three commercial stages (Physik Instrumente) mounted in series, with associated positional sensors.\supercite{lavely_SPS2020} The system provides a total of 13 actuated degrees of freedom (DOF), of which seven  are regularly used during tomographic data collection. For larger scale (cm) motion, a 6-DOF hexapod stage is mounted at the base of the SPS. A high-precision 6-axis flexure stage is mounted atop the hexapod and provides positional accuracy of a few nm across a translational range of 200~$\mu$m. The flexure stage utilizes piezoelectric actuators mounted to flexure bearings, eliminating backlash associated with screws, gears, and roller bearings. A single-axis rotation stage is mounted atop the flexure stage and utilizes a ratcheting piezoelectric actuator with an incremental encoder. Each stage contains internal sensors for positional control, while external interferometric and capacitative distance sensors provide additional, global position sensing capabilities. The entire SPS is shown in Fig. \ref{fig:stage}A.

A custom sample holder assembly is attached to the output of the rotation stage (Fig. \ref{fig:stage}B). The graphite puck with the mounted IC is placed into a circular recess on the sample holder and held in place with a face plate on the front side and spring-loaded brass clips on the back. The entire sample holder is locked on the sample receiver base by a spring-loaded locking mechanism. The mechanism consists of a hook attached to the sample receiver base which rotates downward against a rod attached to the underside of the sample holder, securing the sample holder against the sample receiver base (Fig. \ref{fig:stage}C).


\section{Positional Drift} \label{sec:drift}
During data collection, the electron beam dwells at a specific scan location on the target layer for 45 to 60 seconds, depending on the scan being collected and the rotation angle. The positional drift of the electron beam on the target surface and of the sample stage during this dwell period impacts the achievable spatial resolution by effectively blurring the spot size by a magnitude proportional to the amount drifted. Therefore, the position of the electron beam and stage must remain stable to within the resolution goal of the measurement, with any positional drift resulting in blurring of the final tomographic reconstruction. In addition to instabilities inherent to the stage and electron column, interference from ambient magnetic fields can cause instability in the electron beam. This can lead to unwanted noise in SEM images and poor positional control during tomographic data collection. An electromagnetic interference (EMI) cancellation system (Spicer Consulting) is used to mitigate the effects of ambient fields around MINT.

\begin{figure}[!b]
\centering
\includegraphics[scale=0.6]{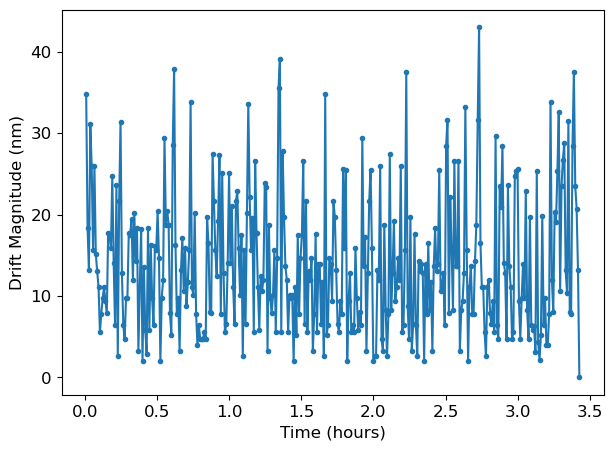}
\caption{Positional drift magnitude per minute over a 3.5 hour period. The positional stability of the SEM beam and stage is within the limit needed to perform nanoscale tomography.}
\label{fig:drift}
\end{figure}

The W fiducials were used to characterize the positional stability of the electron beam and sample stage. During a drift characterization test, an SEM image of the W fiducial is taken prior to any stage movement. The flexure stage is then moved to the intended dwell location for one minute. After the dwell period, the flexure stage is moved back to the previous W fiducial position and another SEM image is acquired. The images are compared and aligned, with the movement needed to realign the fiducial in each image corresponding to the amount of positional drift during the dwell. This process is repeated for a number of dwells, with each post-dwell fiducial image being compared to the fiducial image taken after the dwell prior. Fig. \ref{fig:drift} shows a plot of the positional drift magnitude per minute over a 3.5 hour period. Over all data sets used for tomographic reconstruction, the mean drift per dwell was 15 nm, at dwell times of approximately 45 seconds.


\section{Contrast}
\begin{figure}[!h]
\centering
\includegraphics[scale=0.6]{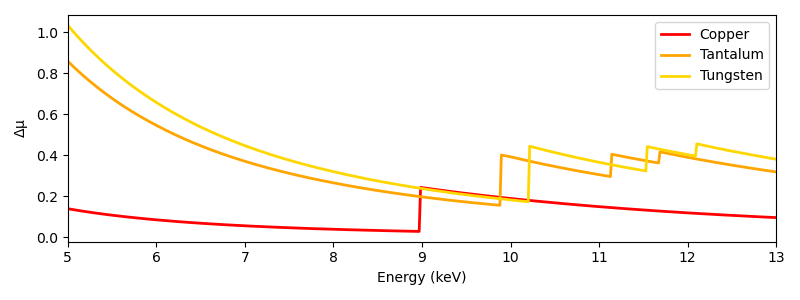}
\caption{$\Delta\mu$ between a given material and SiO$_2$ for Cu, Ta, and W. Cu is the predominant wiring and via material in the sample IC, while Ta and W are candidate materials for the first via layer. The first via layer becomes significantly clearer when including the 5.4-6.4 keV band, indicating a non-Cu material may be present.}
\label{fig:contrast}
\end{figure}

We defined the contrast between Cu and SiO$_2$ as the difference in attenuation $\Delta\mu$. The contrast is a function of the incident photon energy and the material being imaged, and its value at a certain energy is representative of how well photons at that energy differentiate Cu features from SiO$_2$. This calculation can be performed for any candidate material, and can serve as an indicator for which energy bands will be most useful for reconstructions of a given material. Fig. \ref{fig:contrast} shows $\Delta\mu$ for common IC materials.

\printbibliography